\documentclass[a4paper,12pt,onecolumn]{IEEEtran}
\usepackage{amsmath,amssymb,verbatim}
\usepackage[dvips]{graphicx}
\usepackage{setspace}
\newtheorem{definition}{Definition}[section]
\newtheorem{thm}{Theorem}[section]
\newtheorem{proposition}[thm]{Proposition}

\newtheorem{exam}{Example}[section]

\newtheorem{remark}{Remark}[section]

\newcommand{\CC}{\mathbb{C}}

\newcommand{\Cc}{\mathcal{C}}

\newcommand{\Gc}{\mathcal{G}}

\newcommand{\Z}{{\mathbb Z}}
\newcommand{\R}{{\mathbb R}}
\newcommand{\C}{{\mathbb C}}

\newcommand{\HH}{{\mathbf H}}
\newcommand{\OO}{{\mathcal O}}

\newcommand{\A}{{\mathcal A}}

\newcommand{\Q}{{\mathbb Q}}

\newcommand{\mindet}[1]{\hbox{\rm det}_{min}\left( #1\right)}

\begin{document}

\title{Algebraic Hybrid Satellite-Terrestrial Space-Time Codes for Digital Broadcasting in SFN}
\author{\IEEEauthorblockN{Camilla Hollanti$^*$, \emph{Member, IEEE},  Roope Vehkalahti$^*$, \emph{Member, IEEE}, and Youssef Nasser$^{\ddag}$, \emph{Member, IEEE}}\\
\IEEEauthorblockA{$^*$Department of Mathematics,  FI-20014 University of Turku, Finland\\
%\\ e-mails: \{cajoho,\,roiive\}@utu.fi\\
$^\ddag$American University of Beirut, ECE Department, Bliss Street, Beirut, Lebanon \\
}
}

\maketitle

%*********************************************************************%
%
% ABSTRACT
%
%*********************************************************************%

\begin{abstract}
%%%%%%%%%%%%%
Lately, different methods for broadcasting future digital TV in a single frequency network (SFN) have  been  under an intensive study. To improve the transmission to also cover suburban and rural areas, a hybrid scheme may be used. In hybrid transmission, the signal is transmitted both from a satellite and from a terrestrial site. In 2008,  Y. Nasser \emph{et al.} proposed to use a double layer 3D space-time (ST) code in the hybrid $4\times 2$  MIMO transmission of digital TV. In this paper, alternative codes with simpler structure are proposed for the $4\times 2$ hybrid system, and new codes are constructed for the $3\times 2$ system. The performance of the proposed codes is analyzed through computer simulations, showing a significant improvement over simple repetition schemes. The proposed codes prove in addition to be very robust in the presence of power imbalance between the two sites.

\end{abstract}

%*********************************************************************%
%
% INTRODUCTION
%
%*********************************************************************%

\section{Background}

For many years now, multiple-input multiple-output (MIMO) systems are known to provide increased diversity and reliability in wireless communications. For such systems, simple repetition, spatial multiplexing or space-time/space-frequncy (ST/SF) codes exploiting delay can be used to encode the symbols to be transmitted. Here, we concentrate ST/SF codes, while comparison in simulations is also given against simple repetition systems. 

Especially since the work by Sethuraman \emph{et al.} in 2003 \cite{SRS}, algebraic number fields and division algebras have been a standard tool for constructing efficient and robust space-time codes.  Since hybrid SFN ST schemes correspond to joint MIMO transmission from satellite (SAT) and terrestrial (TER) sites, the same methods can be adopted here as well. On the other hand, we can also take advantage of the theory built upon parallel MIMO systems. Namely, in a multi-frequency network (MFN), some of the handheld devices may be able to receive the TER signal only, and hence for such a recipient we should consider the SAT and TER channels as parallel ones. Some of the codes are constructed with also this perspective in mind, and will thus properly function in both SFN and MFN, where the SAT and TER links transmit in parallel frequencies. 

At the moment, the inclusion of a hybrid transmission model to the digital video broadcasting (DVB) next generation handheld (NGH) standard \cite{DVB} is being investigated, setting forth a strong call for codes suitable for such transmission.

In the literature, it is exceptional to find some works dealing with the joint SAT-TER transmission. In \cite{Nasser3D},  a 3D code, proposed initially to cope with TER signals only, has been adopted to achieve this joint transmission. It has been shown that the 3D code presents always better performance than the conventional Alamouti and Golden code but at the detriment of higher complexity. 

In this paper, our aim is to propose efficient ST/SF codes based on an explicit construction for hybrid SFN transmission. The contribution of this work is multi-fold. First, we review the state-of-the-art algebraic lattice ST/SF codes construction. Second, we define the main properties suitable for such a construction. Third, we propose and analyze efficient codes by an explicit construction through the separation between the SAT and TER components. Some of the presented codes are previously known, now being proposed for a novel use, while some of the codes are new  constructions especially designed for the hybrid transmission. Previously known constructions will be properly cited. The codes are proposed to work either with a total number of  2, 3 or 4 transmitting antennas (Tx). 

This paper is structured as follows. Section 2 describes the transmission model. In section 3, we define lattice space-time codes and define the main properties useful for joint SAT-TER transmission. Basic definitions related to algebraic number fields are also given in this section. In section 4, we describe the explicit code constructions for hybrid SAT-TER in SFN environment. Section 5 presents the performance of the proposed codes as well as a summary of the obtained results. Finally, conclusions are drawn in Section 6.

%-----------------------------------------------------------%
%
%CHANNEL
%
%---------------------------%

\section{Hybrid MIMO transmission model}

In this section, we describe the MIMO transmission model of the ST/SF codes constructed between the antennas of the satellite and terrestrial sites. We assume perfect channel state information at the receiver (CSIR) and with no CSIT. The general form of a MIMO transmission model, independently of the SAT and TER components can be written as: 
$$
Y=HX+N,
$$
where $Y,X,H,N$ are the received, transmitted, channel, and the Gaussian noise matrix, respectively. The ST matrix $X\in M_{n_t}(\mathbb{C})$, while $Y,H,N\in M_{n_r\times n_t}(\mathbb{C})$.

One of the main concerns in this hybrid transmission is the land-mobile-satellite (LMS) channel. In literature, various LMS channel models were proposed based on empirical or statistical models \cite{Nasser3D}. However, the three-state model proposed by Fontan \cite{LMS_Fontan} is widely used within DVB-SH \cite{Nasser3D,DVB-SH} since it is recognized as the most accurate statistical LMS channel model available today, encompassing the widest set of environments and elevation angles. In this model, the direct signal is characterized by a three-state first-order Markov chain where the multi-path component could be in  narrow-band or wide-band conditions. These states are defined by:

\begin{itemize}

\item

S1: line-of-sight (LOS) conditions
\item

S2: moderate shadowing conditions
\item

S3: deep shadowing conditions

\end{itemize}

Markov chain models can be used to describe the LMS direct propagation channel at a given route position (equivalently given time) by mean of two matrices:

\begin{itemize}

\item

state transition probability matrix \textit{P} whose each element \textit{P}\textit{$_{ij}$} represents the probability of change from state \textit{i} to state \textit{j}.
\item

state probability matrix \textit{W} (it could be written as a column vector) where \textit{W}\textit{$_{i}$} is the total probability of being in state \textit{i}.

\end{itemize}

The model makes the simplifying assumption of the existence of the three states of the direct signal when traveling a given route. The channel could be in one state according to the probability matrix \textit{W} and the transition from one state to another state is achieved according to \textit{P}. Typically, each state will last few meters along the traveled distance. In \cite{LMS_Fontan}, a minimum distance state length called L$_{min}$ of 3-5 m was observed in the experimentations at S-band. The reader may refer to \cite{LMS_Fontan,LMS_Loo} for more details. However, we should note that the receiver, when moving, may highly suffer from deep or moderate shadowing and hence the performance of the system will be highly degraded if there is no other SAT or TER component.

\begin{remark}In this submitted version of the paper, we make the typical assumption of transmission over a coherent i.i.d. Rayleigh fading channel. This would not change the conclusions of the work since all proposed codes are designed to deal with shadowing effect of the SAT component. 
 For the final version of this paper, we will use a more realistic channel model based on 3-state Markov chains for the SAT link described above. We already have this model implemented, so this is nothing but a matter of waiting for the simulation results to get ready.
\end{remark}

%*********************************************************************%
%
% LATTICES
%
%*********************************************************************%
\section{Lattice codes and number fields}
In this section, we introduce the notion of a lattice space-time code, and provide a very short introduction on number fields. 

\begin{definition}\label{def:stbc}
A {\em space-time} or \emph{space-frequency code} $\Cc$ is a set of $(n\times T)$ complex matrices.
\end{definition}

From now on, we simply talk about ST codes, although all the proposed codes can be equally well used as SF codes.

\begin{definition}
A {\em space-time lattice code} $\Cc \subseteq M_{n\times T}(\C)$ has the form
$$
\Z B_1\oplus \Z B_2 \cdots \oplus \Z B_K,
$$
where the dispersion matrices $B_1,\dots, B_K$ are linearly independent, \emph{i.e.}, form a lattice basis, and $K\leq 2nT$ is
called the \emph{rank} of the lattice. 
\end{definition}
\begin{definition}\label{rate}
With the above notation, the \emph{dimension rate} of  a ST lattice code $\Cc$ is defined as $R_1=\frac K{T}\leq 2n$ $\R$-dimensions per channel use. The (more commonly known) \emph{code rate}  is $R=\frac K{2T}\leq T$ complex information symbols (e.g. QAM) per channel use. 
\end{definition}

For the actual transmission, a finite subset of codewords from $\Cc$ is picked 
by restricting the integral coefficients to some set $\Gc$, \emph{e.g.} to a PAM alphabet. Here, we will only consider  space-time lattice codes and may call
them space-time codes for short.

In SFN, the hybrid code construction follows the guidelines of joint MIMO code design. Both the SAT and the TER site only transmit global information, i.e., signals from both sites contain essentially the same data. For such a joint MIMO transmission, the primary design criterion  for a ST code $\Cc$ is known to be \emph{full diversity} \cite{TSC}, meaning for a  lattice code that
$$
\det(X^\dag X)\neq 0
$$
for all $ {\bf 0}\neq X\in \Cc$. Here $X^\dag$ denotes the conjugate transpose of $X$.
Once achieved,  the next criterion tells us to \emph{maximize the minimum determinant}  \cite{TSC} of the code:
\begin{definition}\label{mindet}
The {\em minimum determinant} $\mindet{\Cc}$ of a space-time  code
$\Cc \subset M_{n_t\times T}(\C)$ is defined to be
\[
\mindet{\Cc}=\inf_{X\neq{\bf 0}}|\det(X^\dag X)|,~ X \in\Cc.
\]
The \emph{normalized minimum determinant} corresponding to unit energy is denoted by $\delta(\Cc)$.
\end{definition}
\begin{definition}\label{NVD}
If further the minimum determinant of the lattice is non-zero,  it is said that the code
has a \emph{non-vanishing determinant} (NVD). This means that the minimum determinant can be lower bounded by a  positive constant independently of the constellation size. In particular, NVD implies full diversity.
\end{definition}

%As DVB-NGH transmission often operates at very low and even negative SNRs, it is not necessarily very severe if the above criteria are not satisfied. We will construct codes that satisfy these criteria as well as codes that do not.

For a multi-frequency network (MFN) employing multiple radio frequencies, the criteria change. There, we should ideally have NVD for 
\begin{equation}\label{PNVD}
\mindet{X_S^\dag X_S}\mindet{X_T^\dag X_T}
\end{equation}
instead of the whole matrix 
$$
X=\left(\begin{array}{c}
X_S\\ \hline
X_T
\end{array}\right),
$$
where $X_S$ describes the  transmission from the SAT site, and $X_T$ the transmission from the TER site. This can be seen by transforming the parallel channel into an equivalent channel, which corresponds to the space-time matrix
$$
X'=\left(\begin{array}{cc}
X_S&0\\ \hline
0& X_T
\end{array}\right).
$$
It is then obvious that we have to consider the product of the determinants instead of the joint determinant (see also \cite{TavVis}).

We call this kind of NVD \eqref{PNVD} \emph{parallel NVD}, and the usual NVD (cf. Def. \eqref{NVD}) \emph{joint NVD} or shortly NVD.

 The notion of relative field norm is used in some of the proposed constructions. Let us shortly recall it  to ease the reading.

Let $E/F$ be a Galois  extension with a Galois group  $\mathrm{Gal}(E/F)=\{\sigma_1,\ldots,\sigma_n\}$, where $n=[E:F]$.
\begin{definition}\label{norm}
Let $a\in E$. The relative field norm of $N_{E/F}:E\rightarrow F$ is defined as the product of the algebraic conjugates:
$$
N_{E/F}(a)=\sigma_1(a)\cdots\sigma_n(a).
$$
We may abbreviate $N(a)=N_{E/F}(a)$, if there is no danger of confusion.
\end{definition}

We denote the ring of algebraic integers of a field $F$ by $\OO_F$. According to basic algebra, we have the following.
\begin{proposition}\label{norminbasefield}
If $a\in\OO_E\setminus\{0\}$, then 
$
N_{E/F}\in \OO_F\setminus\{0\}.
$

In particular, if $F=\Q$ or $F=\Q(\sqrt{-m})$ with $m$ a square free integer, then 
$
N_{E/F}(a)\geq 1
$
for any $a\in\OO_E$.
\end{proposition}

%----------------------------------------------%
%
%HYBRID
%
%------------------------------------------------%

%%%%%%%%%%%%%%%%%%%%%
%%%%%%%%%%%%%%%%%%%%%%
  
\begin{exam} The very first ST code is the Alamouti \cite{Alam} code 
 \begin{equation}\label{alam} X=\left(\begin{array}{c}x_S\\ x_T\end{array}\right)=\left(\begin{array}{cc} c_1&-c_2^*\\ c_2&c_1^*\end{array}\right),
 \end{equation}
  where $c_i\in \CC$ are usually chosen to be QAM symbols $\in\Z[j],\,j=\sqrt{-1}$, and $c^*$ denotes the complex conjugate of $c$. Note that $|c|\geq 1$ for all $c\in\Z[j]$
  
  From an algebraic point of view, the Alamouti code is the matrix representation of a Hamiltonian \emph{quaternion} 
  $$
  q=c_1+uc_2\in \HH,
  $$
  where $u^2=-1$, $ju=-uj$, and $\HH$ denotes the set of Hamiltonian quaternions
  $$
  \HH=\{a+jb+uc+ujd\,|\, a,b,c,d\in\R\}=\C\oplus u\C.
  $$
   Full diversity is already implied by the fact that the quaternions form a division algebra.
   
  In a hybrid transmission, the first row of the Alamouti code is transmitted from a satellite antenna, and the second row from a terrestrial antenna. The transpose of the matrix would correspond to the transmission of plain QAM symbols from the satellite, meaning that the satellite need not be enable to encode. 
  
  The Alamouti code satisfies both the SFN and MFN design criteria, since we have 
  $$
  \det(X^\dag X)\geq 1
  $$
  and 
  $$
  \det(x_S^\dag x_S)\det(x_T^\dag x_T)\geq 1
  $$
  for all $c_1,c_2\in\Z[j], c_1c_2\neq 0$.
  \end{exam}

\begin{exam}
An extension to 4 Tx antennas is provided by the quasi-orthogonal code by Jafarkhani \cite{Jaf_quasiorto},
$$
X=\left(\begin{array}{cc} X_1 & -X_2^*\\
X_2 & X_1^*\\
\end{array}\right),
$$
where $X_1=X_1(c_1,c_2)$ and $X_2=X_2(c_3,c_4)$ are Alamouti blocks and $X^*$ denotes the conjugate of $X$. This code, however, does not have full diversity. 
\end{exam}
%******************************************************************************%
%
% CONSTRUCTIONS
%
%*****************************************************************************%

\section{Explicit code constructions for hybrid SFN}\label{explicit}

\subsection{Codes for 2 SAT and 2 TER Tx antennas}

The simplest option for rate one 4 Tx transmission is the repetition of  the same QAM symbol from the different antennas. The next option that comes to mind is perhaps to use a double-Alamouti code \cite{Double-Alamouti}, i.e., the SAT site and the TER site are both transmitting the same Alamouti block. Such a code admits both joint and parallel NVD, but it is known that simple repetition is not usually the best way to do things. Hence, we propose a few  alternative codes to be used instead. 

If we want to have simple Alamouti encoding at the SAT site while avoiding repetition, we can use the code
$$X=\left(
\begin{array} {r}     X_S\\ \hline 
X_T
\end{array}\right)=
\left(
\begin{array} {cc}            
a  & -b^*\\
b  & a^*\\ \hline
\sigma(a)  & -b^*\\
b & \sigma(a)^*\\
\end{array}\right),$$
where $a\in\Z(\zeta_8),\, b\in\Z[j]$, and  $\sigma: \zeta_8\mapsto -\zeta_8$ is the generator of the cyclic Galois group of the field extension $\Q(\zeta_8)/\Q(j)$. The code has an ``intermediate'' rate of $R=3/2$ QAM symbols per channel use.

In a recent evaluation of MIMO performance, it has turned out that delay is often beneficial when there is imbalance between powers received from the two sites. Hence, we propose yet another rate-one code with higher delay. The following code $L_2$ \cite{HLL} is similar to Jafarkhani's  quasi-orthogonal code \cite{Jaf_quasiorto}, except that the proposed code has full diversity and hence better (non-vanishing) coding gain. This code is a baseline candidate for the DVB-NGH standard \cite{DVB-CFT,DVB}.  Due to its promising performance in the joint $4\times 2$ MIMO transmission of digital TV, we see it as a highly suitable candidate for the hybrid transmission as well. 

The encoding matrix is
$$
X_{L_2}=\left(\begin{array}{c}
X_S\\
X_T\\
\end{array}\right)
=
\left(\begin{array}{rrrr}
c_1 & j c_2 & -c_3^* & -c_4^*\\
c_2 & c_1 & jc_4^* & -c_3^*\\ \hline
c_3 & jc_4 & c_1^* & c_2^*\\
c_4 & c_3 & -j c_2^* & c_1^*\\
\end{array}\right),
$$
where $j=\sqrt{-1}$,  $c_i$ are complex integers (e.g. QAM symbols), and $c^*$  denotes the complex conjugate of $c$. The construction is based on the division algebra $\A=\Q(\zeta_8)\oplus u\Q(\zeta_8)\subseteq \HH$, where again $u^2=-1,\, ju=-uj.$

Assuming there are two satellite antennas and two terrestrial antennas, the first two rows are transmitted by satellite, and the last two by terrestrial antennas. The block structure of the code provides robustness in the case of deep shadowing, and the low rate is advantageous in the presence of high correlation. The code has both joint and parallel NVD. See \cite{HLL} for more details.
 
One may also use this code with 1 SAT antenna and 3 TER antennas in an obvious way. If transposed, the satellite does not  even have to be able to encode, as it is simply transmitting the QAM symbols $c_1,c_2,c_3,c_4$.

%%%%%%%%%
%
% C1
%
%%%%%%%%%%%
For a rate-two transmission, one can again use double-Alamouti to avoid simple repetition of spatial multiplexing, but now the SAT and the TER site transmit independent Alamouti blocks. This however is problematic in SFN, since we are not supposed to have any local data. This can be overcome by switching the 2nd and 3rd row, i.e., by using the slightly modified code
$$
X=\left(\begin{array} {r}     X_S\\ \hline 
X_T
\end{array}\right)=
\left(
\begin{array} {cc}  
a  & -b^*\\
c & -d^* \\ \hline
b  & a^*\\
d & c^*\\
\end{array}\right),$$
where $a,b,c,d\in \Z[j]$, e.g. QAM symbols. The code will have joint NVD, but not parallel NVD.

We propose yet another rate-two code with higher delay to aid the performance in an imbalanced scenario. 
Let  $\zeta=e^{2\pi j/5}$ and  $r=(8/9)^{1/4}$. 
 The encoding matrix is 
$X_{\Cc_1}=M(x_0,x_1,x_2,x_3)$
$$=
\begin{pmatrix}
x_0&-r^2 (x_{1})^*&-r^3\sigma(x_3)&-r\sigma(x_2)^*\\
r^2x_1&(x_0)^*&r\sigma(x_2)&-r^3\sigma(x_3)^*\\ \hline
r x_2&-r^3(x_{3})^*&\sigma(x_0)&-r^2\sigma(x_1)^*\\
r^3 x_{3}&r (x_{2})^*&r^2\sigma(x_{1})&\sigma(x_0)^*\\
\end{pmatrix},
$$
where
$
x_i=x_i(y_1,y_2,y_3,y_4) $$ $$=y_1(1-\zeta) +y_{2}(\zeta-\zeta^2) +y_{3}(\zeta^2-\zeta^3) + y_4(\zeta^3-\zeta^4) 
$
and
$
\sigma(x_i)=y_1(1-\zeta^3)+y_2(\zeta^3-\zeta)+y_3(\zeta-\zeta^4)+y_4(\zeta^4-\zeta^2).
$
This code is based on a cyclic division algebra and is also a candidate for the DVB-NGH standard \cite{DVB-CFT,DVB}. See \cite{FDMIDOsubmission} for more details.

 This code can be used similarly to $L_2$, but provides higher rate $R=2$ (cf. Def. \ref{rate}). The code has joint NVD, but not necessarily parallel NVD. The block structure of the code is very similar to the 3D code \cite{Nasser3D}, and both codes have similar (relatively high) complexities.
 
 Simulations for the rate-two codes will be provided in the final version of the paper.

%%%%%%%%%%
%
%Ideas for new codes
%
%%%%%%%%%%%%%%%%%%%

\subsection{Codes for 1 SAT and 2 TER Tx antennas}

As a simple rate-one construction for 3 Tx antennas, we propose to use a simple combination of QAM and Alamouti transmission:

$$X=\left(
\begin{array} {r}     X_S\\ \hline 
X_T
\end{array}\right)=
\left(
\begin{array} {cc}  a &      b\\         \hline          
a  & -b^*\\
b  & a^*
\end{array}\right),$$

where $a,b\in\Z[j],\, ab\neq 0.$ This code satisfies both the joint and parallel NVD. Namely, a straightforward calculation gives us
$$
\det(X^\dag X)\geq 2
$$
and 
$$
\det(X_S^\dag X_S)\det(X_T^\dag X_T)\geq 1.
$$

Next, we describe a code with a higher rate $R=3/2$, while maintaining joint NVD. To this end, let $a\in\Q(\zeta_8)$, and let $\sigma: \zeta_8\mapsto -\zeta_8$ be the generator of the cyclic Galois group of the field extension $\Q(\zeta_8)/\Q(j)$. We propose to use the encoding matrix 
$$X_{L_3}=\left(
\begin{array} {r}     X_S\\ \hline 
X_T
\end{array}\right)=
\left(
\begin{array} {cc}  a &      b\\         \hline          
\sigma(a)  & -b^*\\
b  & \sigma(a)^*
\end{array}\right),$$
where $ b\in\Z[j]$. We baptize this code as $L_3$.

We have that $a\sigma(a)=N(a)\in\Z[j]$ (cf. Def. \ref{norm}) and $b\in \Z[j]$ so $|a\sigma(a)|, |b|\geq 1$ (cf. Prop. \ref{norminbasefield}). Hence, when $ab\neq 0$,
\begin{eqnarray*}
&& \det( X_SX_S^{\dagger}) \det(X_TX_T^{\dagger})\\
&=& \det( X_SX_S^{\dagger}) |\det(X_T)|^2\\
&=& (|a|^2+|b|^2)(|\sigma(a)|^2+|b|^2)^2\geq 0,
\end{eqnarray*}
and
\begin{eqnarray*} 
\det(X^\dag X)&=&|a|^2+|b|^2+|N(a)|^2+\\ &+&3|b|^2|\sigma(a)|^2+|\sigma(a)|^4+2|b|^4\\
&=&\left\{\begin{array}{rl} |N(a)|^2+|\sigma(a)|^2\geq 1, & b=0,\\
2|b|^4\geq 2, & a=0,\\
|b|^2(|a|^2+3|\sigma(a)|^2+2|b|^2)+& \\+|N(a)|^2+|\sigma(a)|^4\geq 3, & ab\neq 0,
\end{array}\right. 
\end{eqnarray*}
that is, the code has joint NVD and full diversity in the parallel channel.

\begin{comment}
For MFN with some local info from SAT:
$$X=\left(
\begin{array} {r}     X_S\\ \hline 
X_T
\end{array}\right)=
\left(
\begin{array} {cc}  a &      b\\         \hline          
\sigma(a)  & -s^2(a)^*\\
s^2(a)  & \sigma(a)^*
\end{array}\right),$$
\end{comment}

\begin{remark}
In order to achieve NVD in SFN, one may use any space-time code with NVD, e.g. the cyclic division algebra based full-rate Perfect codes \cite{BORV} or the maximal order codes \cite{HLL,HLRV2}. To enable reception with fewer antennas, a full-rate code can always be punctured \cite{spm}. For faster decodability, one may consider the codes proposed in \cite{FDMIDOsubmission}.
\end{remark}

\section{Simulation settings and results}

The aim of this section is to analyze through simulations some of the  codes proposed in the previous sections and to compare with each other and the repetition codes. The channel is assumed to be i.i.d frequency Rayleigh fading. An update to LMS channel is promised by the final submission, for at this moment we are still waiting for the simulations to get ready.  

The OFDM parameters are given in Table \ref{table1}. The spectral efficiencies 2, and 4 b/s/Hz are obtained for different ST schemes as shown in Table \ref{table2}. In all simulations, we assume that two Rx are used by the terminal. 
\begin{table}
\caption{Simulation Parameters}
\label{table1}\begin{center}
\begin{tabular}{|cc|}
\hline

\textbf{Simulation parameter}& \textbf{Value}\\

FFT size &

2048 sub-carriers\\

Sampling Frequency (\textit{f}\textit{$_{s}$}=1/\textit{T}\textit{$_{s}$}) &

9.14 MHz\\

Rate \textit{R}\textit{$_{c}$} of convolutional code &

1/2, 2/3, 3/4\\

Polynomial code generator &

(133,171)$_{o}$\\

Channel estimation&

perfect \\

Constellation (CS)&

QPSK, 16-QAM, 64-QAM\\

Spectral Efficiencies&

\textit{$\eta$}=2 and $\eta$=4 bpcu\\

\hline
\end{tabular}
\end{center}\end{table}

\begin{table}
\caption{Different MIMO Schemes and Efficiencies}
\label{table2}\begin{center}
\begin{tabular}{|c|cccc|}
\hline

\textbf{Spectral Efficiency}&

\textbf{ST scheme} &

\textbf{ST rate }$R$ &

\textbf{Constellation}&

\textit{R}\textit{$_{c}$}\\
\hline
\textit{$\eta$}=2 bpcu &

Alamouti &

1 &

16-QAM &

1/2 \\

&
MISO&

1 &

16-QAM &

1/2 \\

&

$L_3$ &

3/2 &

QPSK &

2/3\\

&

D-Alamouti &

1 &

16-QAM &

1/2\\
&
$L_2$  &

1 &

16-QAM &

1/2\\
\hline\hline
\textit{$\eta$}=4 bpcu &

Alamouti &

1 &

64-QAM &

2/3\\

&
MISO &

1 &

64-QAM &

2/3\\

&

$L_3$&

3/2 &

16-QAM &

2/3\\

&

D-Alamouti &

1 &

64-QAM &

2/3\\

&
$L_2$ code&

1&

64-QAM &

2/3\\

\hline
\end{tabular}
\end{center}\end{table}

\begin{figure}%[htp!]
\begin{center}
\includegraphics[width=13cm]{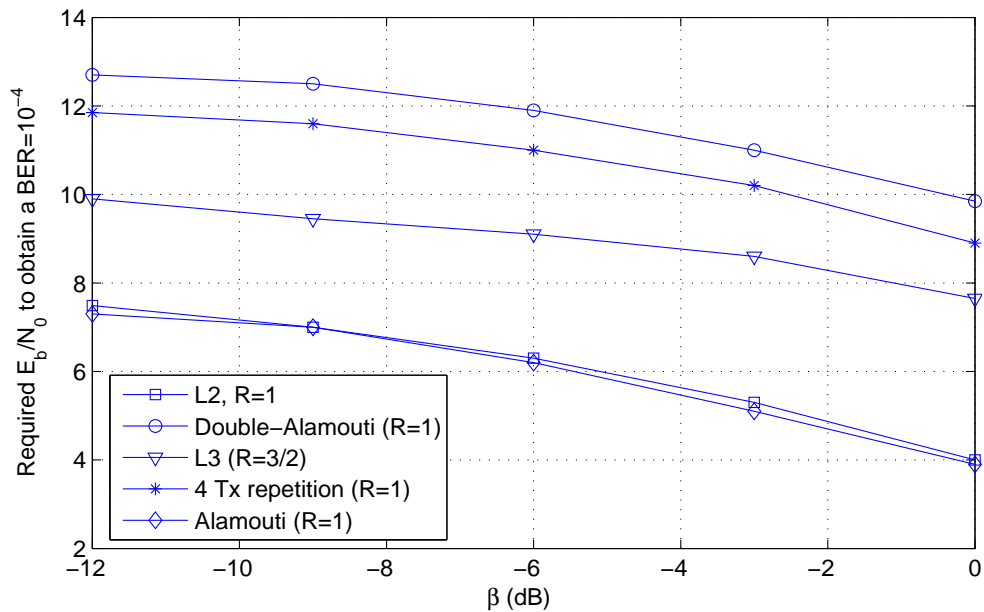}
\end{center}
\caption{Required $E_{b}/N_{0}$ for BER $=10^{-4}$ at 2 bpcu with power imbalance factor $\beta$ ranging from $-12\ldots 0$ dB.}
\label{fig1}
\end{figure}

\begin{figure}%[htp!]
\begin{center}
\includegraphics[width=13cm]{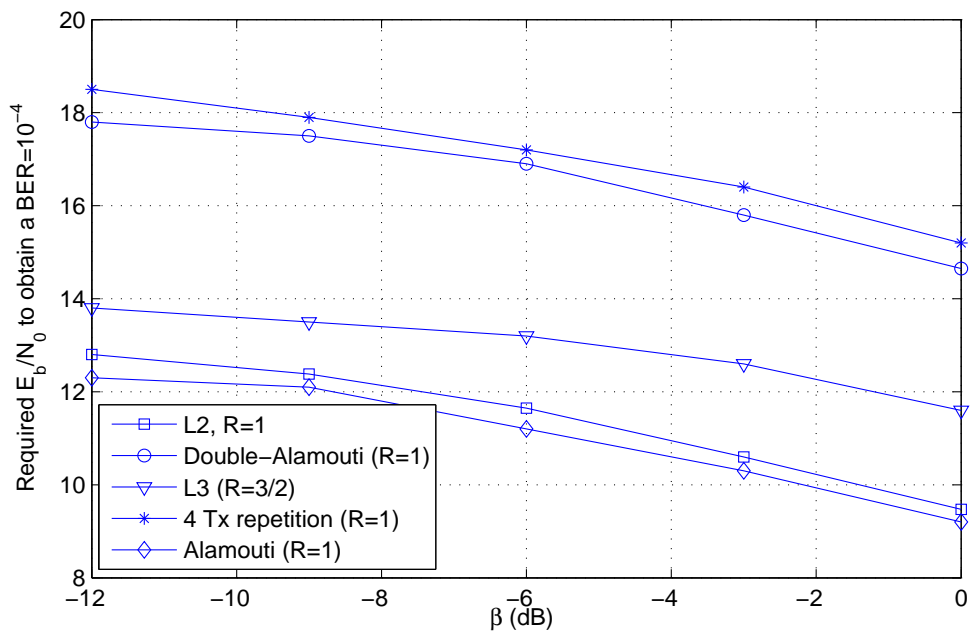}
\end{center}
\caption{Required $E_{b}/N_{0}$ for BER $=10^{-4}$ at 4 bpcu with power imbalance factor $\beta$ ranging from $-12\ldots 0$ dB.}
\label{fig2}
\end{figure}

Figure \ref{fig1} and Figure \ref{fig2} present the required $E_b/N_0$ to obtain a BER $=10^{-4}$ with respect to the power imbalance factor $\beta$, for $\eta=2$ bpcu and $\eta=4$ bpcu respectively, and for the ST codes given in Table \ref{table2}. The factor $\beta$ represents the power difference in dB between the signals' powers received from different Tx antennas. In 4 Tx antennas scenario, we assume that the signal received from 2 Tx antennas is normalized to 0 dB while the signal received from the other Tx has a power equal to $\beta$ dB. For the $L_3$ code, and for fairness purposes, we assume that the signal power received from 2 Tx antennas is normalized to 0 dB while the signal received from the third antenna has a power  normalized to $\beta$ dB. In both figures, it is easily shown that the 2 Tx Alamouti code and the 4 Tx quasi-orthogonal $L_2$ code have almost the same performance. This result is very important since in a configuration where 4 Tx are required for transmission, 2 SAT and 2 TER, for instance, we can ensure the same performance is obtained as with the simple Alamouti code. For hybrid transmission with 3 sites, like in the 3Tx TER SFN or hybrid 1 SAT and 2 TER, the $L_3$ code presents acceptable performance when it is compared to other schemes. However, as could be expected, the rate-one double-Alamouti and the simple repetition (4 Tx) code require higher $E_b/N_0$ to obtain the same performance.


%*********************************************************************%
%
% CONCLUSION
%
%*********************************************************************%

\section{Conclusions}

In this paper, algebraic hybrid SAT-TER ST codes for digital broadcasting have been proposed. An explicit construction scheme has been investigated by the separation between the SAT and TER components. The proposed codes deal with a transmission using 3 or 4 transmitters, while comparing to 2 Tx Alamouti code and 4 Tx repetition codes. We have shown that the $L_2$ (comparable to Alamouti) and $L_3$ codes present the best performance in Rayleigh fading channel, significantly outperforming the repetition schemes regardless of the power imbalance. Additional results with an LMS channel will be given in the final version of the paper.

\label{sec:conc}

%*********************************************************************%
%
% ACK
%
%*********************************************************************%

\section*{Acknowledgments}
The research of C. Hollanti and R. Vehkalahti is supported by the Emil Aaltonen
Foundation's Young Researcher's Project, and by the Academy of Finland 
grant \#131745.

%******************************************************************************%
%
% BIBLIO
%
%******************************************************************************%

%\bibliographystyle{ieee}	% (uses file "plain.bst")
%\bibliography{myrefs_MIDO-IT}		% expects file "myrefs.bib

\end{document}